\documentclass[%
reprint,
superscriptaddress,
amsmath,amssymb,
aps,
pre,
nofootinbib,
nolongbibliography
]{revtex4-2}

\usepackage[usenames,dvipsnames]{xcolor}
\usepackage{amsthm}
\usepackage{amssymb}
\usepackage{mathtools}

\usepackage{diagbox}

\usepackage{graphicx}
\usepackage{dcolumn}
\usepackage{bm}
\usepackage[unicode, colorlinks, urlcolor=blue, linkcolor=blue, pagecolor=blue, citecolor=blue]{hyperref} 
\usepackage[T2A]{fontenc}
\usepackage[utf8]{inputenc}
\usepackage[russian,english]{babel}
\usepackage{comment}
\addto\captionsenglish{}
\usepackage{bigstrut}
\usepackage{float}

\let\oldcfrac\cfrac
\renewcommand{\cfrac}[2]{\oldcfrac{#1}{#2}\,}
\usepackage{comment}
\usepackage{hhline}
\DeclareMathOperator\erf{erf}
\DeclareMathOperator\erfc{erfc}
\begin{document}

	%
	\title{Pressure of Coulomb systems with volume-dependent long-range potentials}
	
	\newcommand\JIHT{Joint Institute for High Temperatures, Izhorskaya 13 Bldg 2, Moscow 125412, Russia}
	\newcommand\MIPT{Moscow Institute of Physics and Technology, Institutskiy Pereulok 9, Dolgoprudny, Moscow Region, 141701, Russia}

	\author{A. S. Onegin}
	\affiliation{\JIHT}
	\affiliation{\MIPT}
	
	\author{G. S. Demyanov}
	\affiliation{\JIHT}
	\affiliation{\MIPT}
	
	\author{P. R. Levashov}
	\affiliation{\JIHT}
	\affiliation{\MIPT}
	
	\date{\today}
	
	\begin{abstract}
		In this work, we consider the pressure of Coulomb systems, in which particles interact via a volume-dependent potential
		(in particular, the Ewald potential).
		We confirm that the expression for virial pressure should be corrected in this case. We show that the corrected virial pressure coincides with the formula obtained by differentiation of free energy if the potential energy is a homogeneous function of particle coordinates and a cell length. As a consequence, we find out that the expression for pressure in the recent paper by J. Liang \textit{et al.} [\href{https://doi.org/10.1063/5.0107140}{J. Chem. Phys. \textbf{157}, 144102 (2022)}] is incorrect.
	\end{abstract}
	
	\maketitle
	
	\section{Introduction}
	The importance of Coulomb systems in statistical physics is undoubtful due to their numerous applications including electrolytes \cite{Attard:CS:2002}, molten salts \cite{Roper:MoltenSalt:2022}, liquid metals \cite{Bo:APX:2018, Smirnov:JAppPhys:2023}, astrophysics \cite{PotekhinAddendum:PRE:2009, Potekhin:AA:2013}, and plasmas \cite{Fortov:NPP:2006,Redmer:CPP:2010}. Despite the simplicity of the Coulomb potential, the variety of physical phenomena in systems of charged particles is truly remarkable. 

	Many interesting features of Coulomb systems arise from the long-range nature of the potential. These include stability and thermodynamic limit \cite{Lieb:RMP:1976,Baus:PR:1980} problems, thermodynamic and dielectric  properties of dense plasma \cite{Militzer:PRE:2021, Moldabekov:JChemPhys:2023}, and atomistic simulation \cite{Dornheim:PR:2018,DemyanovOCP:PRE:2022}. The latter faces many difficulties because the slow decrease of the potential involves many particles in the calculation of the interaction energy.

	To reduce the number of particles in a computational cell, periodic boundary conditions (PBC) are traditionally used \cite{AllTID,Frenkel:MD:2002, Filinov:PRB:2023}. In the case of Coulomb systems, however, it is necessary to sum up the interactions not only between particles in the main cell, but also between all their periodic images \cite{Rapaport:MD:2004}.

	To compute such a conditionally convergent series, the well-known Ewald technique was proposed  \cite{Ewald:1921}, which has been widely used in many applications.
	The result of this technique is an effective Ewald potential between particles only in the main computational cell.
	However, the Ewald potential has a complicated analytical form of a rapidly converging series; so, the calculation of thermodynamic properties for a system with such a potential can be tricky. In addition, this potential explicitly depends not only on the particle coordinates, but also on the length of the cubic simulation cell. 
	
	The pressure of a system of particles can be calculated using a number of techniques. The most obvious is to differentiate the free energy by volume to express the pressure through particle positions.
	However, the pressure is often calculated using the force virial of the system that is readily available \cite{AllTID}. The relationship between the internal virial and the pressure is derived from the virial theorem, which relates all the forces acting on particles to the time-averaged kinetic energy of a system (see Eq.~(2.51) and reasoning below in Ref.~\cite{AllTID}). Note that for some potentials, including the Ewald one, the potential energy can depend on the system volume or, equivalently, on the characteristic length of the computational cell. 
	
	If the potential energy is a homogeneous function of particle coordinates, then the virial and, consequently, the pressure can be expressed through the potential energy \cite{LandauLifshits:V5:2013}. We refer to the relationship between the virial and the pressure as the virial pressure. This relationship is a well-known consequence of the virial theorem\footnote{The virial theorem is sometimes called the relationship between the potential energy and the virial in the case of a homogeneous potential \cite{LandauLifshits:V5:2013}.} and takes a simple form for the Coulomb potential.
	However, the Ewald potential and angular averaged Ewald potential (AAEP) \cite{Yakub:2003,Demyanov:JPhysA:2022,DemyanovOCP:PRE:2022} are not homogeneous in coordinates due to an explicit dependence on the cell length. Therefore, the validity of the virial pressure expression is doubtful. In this case, the expression for the virial pressure should be corrected to be consistent with statistical mechanics \cite{Louwerse2006,Thompson:JCP:2009}.

	The virial approach for pressure calculation has been used earlier in simulations of a one-component Coulomb system (OCCS) or one-component plasma (OCP), including in the very first work \cite{BST:1966}. This classic paper considers an OCP with the Ewald potential and uses the virial definition of pressure. However, the validity of this definition is not obvious for the Ewald potential.  


	An explicit dependence of potential on volume does not only arise in atomistic simulations with PBC. In average atom models, where an atom or ion is confined within a spherical cell \cite{Nikiforov:book:2005}, the self-consistent
	electrostatic potential depends  on both temperature and volume. This dependence originates significant difficulties
	in the calculation of the second derivatives of the free energy for the Thomas--Fermi model \cite{Shemyakin:JPA:2010,Shemyakin:CPC:2019}.
	The assumed thermodynamic inconsistency in the Liberman model \cite{liberman:PRB:1979,Ovechkin:HEDP:2014}
	is probably caused by the neglect
	of the self-consistent potential volume dependence when the virial theorem is applied. 
	
	In this paper, we consider the problem of pressure calculation for a system in which particles interact via a
	volume-dependent potential. In particular, we analyze both the Ewald potential and the AAEP and show that the virial theorem and the differentiation of free energy lead to the same result for the pressure. 
	We also show that the expression for the pressure in the recent paper \cite{Liang_2022} is incorrect. We derive the correct expression in which the pressure is completely defined by the potential energy.

	Thus, we consider a classical (nondegenerate) two-component Coulomb system (TCCS), also called a two-component plasma (TCP), and a one-component Coulomb system (OCCS), also called a one-component plasma (OCP). Although plasma is a more restricted concept compared to a Coulomb system, its descriptive formalism is the same. We use either the term ``Coulomb system'' or ``plasma'', which are equivalent in this work.

	\section{Theoretical background}
	\label{sec:theory}
	
	In this section we introduce the basic notations for the Coulomb systems under consideration. The pressure is determined via the virial and by differentiating the free energy. Finally, we derive the expression for the pressure for a system with PBC.

	\subsection{Charged particles in a cell}
	In this paper, we consider a cubic (main) cell of  volume $V = L^3$ containing $N$ charged dotted particles. The entire cell is electrically neutral.  
	Electroneutrality can be achieved by compensating the charges of particles (in the case of positively and negatively charged particles, recalled as a TCCS or a TCP) or through a compensating homogeneous background (in the case of an OCCS or OCP).
	These particles interact via a pair potential $\phi(\textbf{r})$. We impose periodic boundary conditions (PBC) on the cell. Thus, each particle has an infinite number of images. Every $i$th particle is located at a point $\textbf{r}_i$ in the main cell and has a momentum $\textbf{p}_i$ and mass $m$.
	
	The full energy, $H = K + U$, of particles in the cell \emph{for a given configuration} consists of the kinetic, $K$, and the potential, $U$, energy. The kinetic energy is calculated as a sum of kinetic energies of all particles:
	\begin{equation}
		K = \sum_{i = 1}^N\cfrac{\textbf{p}_i^2}{2m}.
	\end{equation}
	Since the images of all particles create some potential at the point $\textbf{r}_i$, the potential energy has the following form (we use Gaussian units):
	\begin{equation}
		\label{eq:fullPotEnergyGeneral}
		U = \cfrac{1}{2} \sum_{i = 1}^Nq_i\sum_{j = 1}^N\sideset{}{'}\sum_{\textbf{n}}q_j\phi(\textbf{r}_{ij} + \textbf{n}L) + \text{const},
	\end{equation}
	where $q_i$ is the charge of the $i$th particle, ${\textbf{r}_{ij} = \textbf{r}_i-\textbf{r}_j}$, and $\textbf{n} = (n_x, n_y, n_z) \in \mathbb{Z}^3$ represents the vector of integers. The prime in Eq.~\eqref{eq:fullPotEnergyGeneral} means that the terms with $\textbf{n}=\textbf{0}$ are omitted if $i = j$ to prevent the interaction of the particle with itself. The interactions between the $i$th particle and all of its images are included in the sum~\eqref{eq:fullPotEnergyGeneral}. The const-term in Eq.~\eqref{eq:fullPotEnergyGeneral} denotes the contribution of a homogeneous background if it is present in the system; otherwise the term is zero.
	
	The probability $w$ of each configuration in the canonical ($NVT$) ensemble at a temperature $T$ is given by:
	\begin{equation}
		\label{eq:partitionFuncDef}
		w = \cfrac{\exp(-\beta H)}{Z}, \quad Z = \cfrac{Q}{N!\Lambda^{3N}},
	\end{equation}
	where $\beta = (k_B T)^{-1}$ is the inverse temperature, $k_B$ denotes the Boltzmann constant, $\Lambda = \Lambda(T) = \sqrt{2\pi\hbar^2\beta/m}$ is the thermal de Broglie wavelength, and $Z$ represents the partition function. Here, $Q$ is the classical configuration integral:
	\begin{equation}
		\label{eq:confInt}
		Q = \int d \textbf{r}_1\ldots d\textbf{r}_N\exp\left(-\beta U\right).
	\end{equation}
	Thus, the full and potential system energy are calculated as an average over the ensemble:
	\begin{equation}
		\!\!\!  E \! = \!
		\tfrac{1}{Z}\!\int\! H e^{-\beta H}d\textbf{r}_1\ldots d \textbf{r}_N d \textbf{p}_1\ldots d \textbf{p}_N \! =\!
		\cfrac{3Nk_B T}{2} + \langle U\rangle,
	\end{equation}
	where the thermodynamic potential energy, $\langle U\rangle$,  is:
	\begin{equation}
		\label{eq:potEnergyTherm}
		\langle U\rangle \equiv \cfrac{1}{Q} \int U e^{-\beta U}d\textbf{r}_1\ldots d \textbf{r}_N.
	\end{equation}
	
	\subsection{Definitions of pressure}

	As we consider the $NVT$-ensemble, it is natural to calculate pressure  as the derivative of the Helmholtz free energy, $F$, with respect to $V$ at constant $T$:
	\begin{equation}
		\beta P_F = - \left(\cfrac{\partial (\beta F)}{\partial V}\right)_T, \quad \beta F = -\ln Z.
	\end{equation}
	Since the denominator of the partition function in Eq.~\eqref{eq:partitionFuncDef} depends only on temperature, 
	the pressure can be obtained by differentiating the configuration integral~\eqref{eq:confInt} with respect to $V$:
	\begin{equation}
		\label{eq:dQ/dV}
		\beta P_F=\left(\frac{\partial \ln Q}{\partial V}\right)_T=\frac{1}{Q}\left(\frac{\partial Q}{\partial V}\right)_T.
	\end{equation}
	
	To calculate the derivative in Eq.~\eqref{eq:dQ/dV}, the length of the main cell is scaled by $\gamma$ times as follows:
	\begin{multline}
		L=\gamma L_0, V=\gamma^3 V_0,\  \textbf{r}_{i}=\gamma \textbf{r}_{i,0}, \\d\mathbf{r}_i = \gamma^3 d\mathbf{r}_{i,0},\ dV=3 V d\gamma/\gamma,
	\end{multline}
	where $\textbf{r}_{i,0}$ are particle coordinates in the cell of the initial volume $V_0$. As a result, the derivative with respect to $V$ is replaced by the derivative with respect to  $\gamma$:
	\begin{multline}
		\left(\frac{\partial Q}{\partial \gamma}\right)_T  =  \int d\textbf{r}_{1,0}...d\textbf{r}_{N,0} \frac{\partial}{\partial \gamma}\left(\gamma^{3N} e^{-\beta U}\right)_T  \\
		= 3N \gamma^{-1}Q + \int d\textbf{r}_{1}...d\textbf{r}_{N}  \left(-\frac{\partial (\beta U)}{\partial \gamma} \right)_T e^{-\beta U} .
	\end{multline}
	Then we substitute the derivative by $\gamma$  in~\eqref{eq:dQ/dV} and obtain the expression for the pressure, $P_F$:
	\begin{multline}
		\label{eq:pvkt}
		\frac{\beta P_F V}{N}= 
		\frac{\gamma}{3  N Q } \left(\frac{\partial Q}{\partial \gamma}\right)_{T}
		=
		1 - \frac{\gamma}{3  N }\left\langle \left(\frac{\partial (\beta U)}{\partial \gamma} \right)_T\right\rangle
		\\=1 - \frac{ V}{N }\left\langle \left(\frac{\partial (\beta U)}{\partial V}\right)_T\right\rangle, 
	\end{multline}
	where $\langle (\ldots)\rangle$ means the statistical averaging (see Eq.~\eqref{eq:potEnergyTherm}). Formula~\eqref{eq:pvkt} coincides with Eq.~(4) in Ref.~\cite{Louwerse2006}.
	The first term in Eq.~\eqref{eq:pvkt} corresponds to the ideal gas contribution, and the second one is the excess pressure due to the non-ideality of the system.

	Another way to determine pressure is based on the virial theorem \cite{Tsai:JChemPhysL:1979}. According to the book of Allen and Tildesley (see Eq.~(2.53) of Sec. 2.4 in Ref.~\cite{AllTID}), there is a relationship between the virial, $W$, and pressure for a classical system: 
	\begin{equation}
		\label{pv:virial}
		\frac{\beta P_W V}{N}  = 1 +  \frac{\langle \beta W\rangle_{\tau}}{3N}, 
	\end{equation}
	where the (internal) virial is:
	\begin{equation}
		\label{virial definition}
		W = \sum_{i=1}^{N} \textbf{r}_i \cdot \textbf{f}_i.
	\end{equation}
	Here, $\textbf{f}_i$ is the force acting on the $i$th particle from the other particles.
	It is also convenient to write the virial in the pair form (see the derivation of Eq. \eqref{eq:virialPairFormDef}  in Appendix~\ref{app:virialPair}):
	\begin{equation}
		\label{eq:virialPairFormDef}
		W = \frac{1}{2} \sum_{i=1}^N \sum_{\substack{j=1\\ j\neq i}}^N \textbf{r}_{ij}\cdot \textbf{f}_{ij},
	\end{equation}
	where $\textbf{f}_{ij}$ is the force acting on the $i$th particle from the $j$th particle.
	
	In Equation~\eqref{pv:virial} the notation $\langle(\ldots)\rangle_{\tau}$ means time averaging. We use the mechanical definition of the virial (see Sec. 3.4 in Ref.~\cite{goldstein:mechanics} or Sec. 3.5 in Ref.~\cite{Park1990}).
	
	\subsection{\label{sec:PBC}Influence of periodic boundary conditions}
	In simulations with PBC, the potential energy in Eq.~\eqref{eq:fullPotEnergyGeneral} explicitly depends on the length of the computational cell, $L$.
	Thus, we can write $U = U(\textbf{r}_1, ..., \textbf{r}_N; L)$. In this case, formula~\eqref{pv:virial} may produce wrong results~\cite{Louwerse2006}, since it is necessary to take into account the variation from both the scaling of $L$ at fixed particle positions, and the scaling of the particle positions. This can be obtained by writing the  derivative of $U(\textbf{r}_1, ..., \textbf{r}_N; L)$ in Eq.~\eqref{eq:pvkt} using the chain rule:
	\begin{multline}
		\label{eq:dU/dV} 
		\left(\frac{\partial U}{\partial V}\right)_T = 
		\left[\sum_{i=1}^N \left(\frac{\partial U}{\partial \textbf{r}_i}\right)_{T, L} 
		\cdot
		\frac{d\textbf{r}_i}{dL} 
		+ 
		\left(\frac{\partial U}{\partial L}\right)_{T, \textbf{r}_i} \right]\frac{dL}{dV} 
		\\= 
		-\frac{1}{3V} \sum_{i=1}^N \textbf{r}_i \cdot \textbf{f}_i
		+ 
		\frac{L}{3V} \left(\frac{\partial U}{\partial L}\right)_{T, \textbf{r}_i}.
	\end{multline}
	Here we use the relation $d\textbf{r}_i/dL = \textbf{r}_i/L$, and the definition of the force acting on the $i$th particle, ${\textbf{f}_i = - \left(\partial U/\partial \textbf{r}_i\right)_{T,L}}$.
	
	The notation $(\partial/\partial L)_{T, \textbf{r}_i}$ means that the derivative is taken only by the parameter $L$, on which the potential energy explicitly depends. The volume dependence of the coordinates is taken into account in the first term of Eq.~\eqref{eq:dU/dV}, $\left(\partial /\partial \textbf{r}_i\right)_{T, L} $.

	Substituting Eq.~\eqref{eq:dU/dV} into formula~\eqref{eq:pvkt}, we obtain the the relationship between the virial and the pressure of the systems whose energy depends on the cell length, $L$:
	\begin{equation}
		\label{correct pv:virial}
		\frac{\beta P_F V}{N} = \left[1 +
		\frac{\left\langle \beta W\right\rangle}{3N}\right]
		-
		\frac{1}{3N}\left\langle L\left(\frac{\partial (\beta U)}{\partial L}\right)_{T, \textbf{r}_i} \right\rangle.
	\end{equation}
	Thus, an explicit dependence of the potential energy on the cell size results in an additional contribution to the pressure. It seems challenging to derive this pressure contribution from the dynamic consideration, in particular directly from the virial theorem. At the same time, there is no reason to believe that the virial theorem and the free energy differentiation should lead to different results for pressure.
	
	Similar expressions were previously obtained  \cite{Louwerse2006,Thompson:JCP:2009} and numerically verified for the Lennard-Jones,  Stillinger-Weber, and ReaxFF potentials. However, no analytical analysis was performed for any potentials. In Section~\ref{sec:pressureCoulomb}, we are going to apply Eq.~\eqref{correct pv:virial} to both the Ewald potential and AAEP.
	
	\section{Ewald summation technique}
	\label{sec:applications}
	
	
	Further in the paper we are going to deal with Coulomb systems. It means that all the particles interact via the Coulomb potential, $\phi(\textbf{r}) = 1/r$, where $r = |\textbf{r}|$.
	
	Considering Coulomb plasma systems, one is interested in thermodynamic properties in the thermodynamic limit \cite{DemyanovOCP:PRE:2022, Caillol:JChemPhys:1999}. This means that the cell size, as well as the number of particles $N$, tend to infinity at a fixed particle density, $N/V$. As the size of the system increases, the influence of boundary conditions decreases.
	
	If the main cell is large enough, we can consider only particles in the main cell, which corresponds to the term with $\textbf{n}=\textbf{0}$ in Eq.~\eqref{eq:fullPotEnergyGeneral}. Thus, the interaction potential $\phi(\textbf{r})$, as well as the potential energy, is independent of $L$:
	\begin{equation}
		U\to U_{\textbf{n}=\textbf{0}} = \tfrac{1}{2}\sum_{i\neq j}q_iq_j\phi(\textbf{r}_{ij})
		\Rightarrow
		\left(\cfrac{\partial U_{\textbf{n}=\textbf{0}}}{\partial L}\right)_{\textbf{r}_i} = 0.
	\end{equation}
	
	In this case, the interaction potential is homogeneous in coordinates; so, the virial pressure is valid for sure. Therefore, the expressions for $P_F$ and $P_W$ (see Eqs.~\eqref{correct pv:virial} and~\eqref{pv:virial}, respectively) have the same form:
	\begin{equation}
		\label{eq:1+U/3N_coulomb}
		\frac{\beta P V}{N} = 1 + 
		\frac{\left\langle \beta U\right\rangle}{3N}, \quad N\to\infty, \  U\to U_{\textbf{n}=\textbf{0}}.
	\end{equation}
	To calculate $W$ from Eq.~\eqref{eq:virialPairFormDef}, the following relations were used:
	\begin{equation}
		\phi(\textbf{r})=\frac{1}{r}, \quad \textbf{r}\cdot\frac{\partial \phi(\textbf{r})}{\partial \textbf{r}} = r\phi^\prime_r = -\frac{1}{r} \Longrightarrow  W = U.
	\end{equation}

	For an OCP, the whole sum in Eq.~\eqref{eq:fullPotEnergyGeneral} was calculated using the Ewald summation method \cite{BST:1966, Hansen:PRA:1973}. Nevertheless, expression~\eqref{eq:1+U/3N_coulomb} was used to obtain OCP pressure even for small computational cells and $N < 100$. It is not clear why Eq.~(\ref{eq:1+U/3N_coulomb}) is valid in this case.
	
	We first review the main results of the Ewald technique in Sec.~\ref{sec:ewaldSummTechnique}, and then show in Sec.~\ref{sec:pressureCoulomb} that Eq.~\eqref{eq:1+U/3N_coulomb} is exact for (classical) Coulomb systems of any $N$.

	\subsection{Ewald potential for one- and two-component Coulomb systems}
	\label{sec:ewaldSummTechnique}
	To take into account the problem of long--range interaction, PBC are imposed on the computational cell and the Ewald procedure is applied.
	To obtain the electrostatic energy, $U$, the Poisson equation with PBC is solved.
	Thus, the sum~\eqref{eq:fullPotEnergyGeneral} can be rewritten in a more convenient form for simulations:
	\begin{equation}
		\label{eq:TCP_full}
		U = U_0 + \cfrac{1}{2}\sum^N_{i = 1}\sum_{\substack{j=1 \\ j\neq i}}^Nq_iq_jv(\textbf{r}_{ij}; L),
	\end{equation} 
	where $v(\textbf{r}; L)$ denotes the Ewald potential, $U_0$ is the constant that accounts for the interaction of each particle with its image, including the interaction between the particles and neutralizing background in the case of an OCP.
	
	To calculate the sum in Eqs. \eqref{eq:TCP_full} and \eqref{eq:aaepEnergy}, we use the minimum image convention. 
	It states that while calculating the interaction between the two particles with numbers $i$ and $j$, one should consider the closest to the $i$th particle image of the $j$th one in the neighboring cells, rather than all possible images of the $j$th particle (see Sec. III of Ref. \cite{BST:1966}).
	
	The Ewald potential follows this principle. Thus, the $i$th particle interacts only with the particles in the main cell or the ones from neighboring periodic images, if they are closer to the $i$th particle. This is achieved because the Ewald potential itself accounts for interactions with an infinite number of periodic images.
	
	OCCS (or OCP) and TCCS (or TCP) are examples of Coulomb systems. Particles in these systems interact through the Coulomb potential. In the OCCS, in addition to particles, there is a uniform neutralizing background. This background can be represented as a continuum of point charges interacting with dotted ions, and the system will remain Coulomb.
	
	We denote the energy of pair interactions as $U_\text{pair}$:
	\begin{equation}
		U_\text{pair} \coloneqq U - U_0.
	\end{equation}
	The pair Ewald potential, $v(\textbf{r}; L) = v_1(\textbf{r}; L) + v_2(\textbf{r}; L)$, in the case of OCCS and TCCS is the same:
	\begin{equation}
		\label{v_1}
		v_1\left(\textbf{r};L\right) = \frac{1}{L}\sum_{\textbf{n}} \frac{\erfc \left(\delta \left|\textbf{r}/ L+ \textbf{n}\right|\right)}{\left|\textbf{r}/ L+ \textbf{n} \right|},
	\end{equation}
	\begin{equation}
		\label{v_2}
		v_2\left(\textbf{r};L\right) = \frac{1}{L}\sum_{\textbf{n} \neq \textbf{0}} \frac{e^{-\pi^2 n^2/\delta^2}}{\pi n^2} \cos \left( 2\pi \textbf{n} \cdot \textbf{r}/L \right),
	\end{equation}
	where $\delta$ is a \emph{dimensionless} parameter. Note that $\delta$ is independent of the cell volume.
	
	Usually, $\sqrt{\pi}$ is substituted for the $\delta$ parameter in the Ewald potential for an OCP~\cite{BST:1966, Hansen:PRA:1973, DemyanovOCP:PRE:2022}.
	Also all ion charges are equal to $Ze$ in an OCP, where $Z$ is the charge number and $e$ is the electron charge.
	
	The constant terms differ for an OCP:
	\begin{equation}
		\label{eq:OCP_Ewald_const}
		U_0 \to U_0^{\mathrm{OCP}} = \frac{\chi - \pi N/\delta^2}{2L} (Ze)^2N , 
	\end{equation}
	and TCP:
	\begin{equation}
		U_0 \to U_0^{\mathrm{TCP}} = \frac{\chi}{2L} \sum_{i=1}^N q_i^2, 
	\end{equation}
	where
	\begin{equation}
		\chi = \sum_{\textbf{n} \neq \textbf{0}} \left[ \frac{\erfc(\delta n)}{n} + \frac{e^{-\pi^2 n^2 / \delta^2}}{\pi n^2}\right] - \frac{2\delta}{\sqrt{\pi}}. 
	\end{equation}
	One can find the derivation of OCP energy in App. \ref{app:OCPenergyDeltaDeriv} and TCP energy in Ref.~\cite{DeLeeuw1980SimulationOE}.
	
	\subsection{Angular-averaged Ewald potential}
	
	
	Many ``Non-Ewald'' methods have been developed to calculate the electrostatic long-ranged interactions of charged systems~\cite{Fukuda2022} to speed up simulations. We now consider the method called ``pre-averaging'' or ``angular--averaging'' of the Ewald potential.
	
	Its idea is to average $v(\textbf{r}; L)$ over the angles to obtain a spherically symmetric angular-averaged Ewald potential (AAEP) \cite{Yakub:2003}. The AAEP behaves like a Coulomb potential at small distances, and approaches zero with zero first derivative at the point $r_m = ( \frac{3}{4\pi})^{1/3}L$. The calculation procedure with AAEP is the same as in traditional simulations with short--range potentials, the cut-off radius in this case is $r_m$. The reader can find the derivation of AAEP and computational details in Refs.~\cite{Yakub:2003,Jha:2010, Demyanov:JPhysA:2022, DemyanovOCP:PRE:2022}.
	
	Then the electrostatic energy of OCCS and TCCS can be presented as:
	\begin{equation}
		\label{eq:aaepEnergy}
		U^{\text{AAEP}} = U_{0,a} + \frac{1}{2}\sum_{i=1}^N \sum_{\substack{j=1 \\ j\neq i}}^{N_{s,i}} q_i q_j v^a(r_{ij}; L),
	\end{equation}	
	where $N_{s,i}$ is the number of neighbors of the $i$th particle in the sphere of  volume $4\pi r_m^3/3 = L^3$.
	The AAEP is given by the following formula: 
	\begin{equation}
		\nonumber
		v^a(r; L) = \begin{cases}
			\frac{1}{r}\left\{1 + \frac{1}{2}\left(\frac{r}{r_m}\right)\left[\left(\frac{r}{r_m}\right)^2-3\right]\right\}, &r < r_m\\
			0, &r \geq r_m.
		\end{cases}
	\end{equation}
	The constant contribution in the case of TCP and OCP is equal to:
	\begin{equation}
		U_{0,a}\to	U_{0,a}^{\mathrm{TCP}} = -\frac{3}{4} \sum_{i=1}^N \frac{q_i^2}{r_m},
	\end{equation}
	and
	\begin{equation}
		U_{0,a}\to	U_{0,a}^{\mathrm{OCP}} = -\frac{3}{20} (Ze)^2 \frac{N(N + 5)}{L},
	\end{equation}
	respectively.
	
	Since both systems are Coulomb, we expect that the pressure of these systems will be related to the potential energy by formula~\eqref{eq:1+U/3N_coulomb}, which obtained from the virial pressure. In the following sections, we show analytically that formulas~\eqref{eq:1+U/3N_coulomb} and~\eqref{correct pv:virial} coincide for the systems described above.
	

	\section{Pressure of Coulomb systems}
	\label{sec:pressureCoulomb}
	
	The simplest way to calculate the pressure of a Coulomb system is to consider the derivative  $\left(\partial (\beta U)/\partial \gamma\right)_{T}$ in Eq. \eqref{eq:pvkt}. 
	Suppose that $U(\textbf{r}_{1},\ldots, \textbf{r}_{N}; L)$ is a homogeneous function of all variables (including $L$) of degree -1:
	\begin{equation}
		U(\gamma\textbf{r}_{1},\ldots, \gamma\textbf{r}_{N}; \gamma L)
		=\cfrac{1}{\gamma} U(\textbf{r}_{1},\ldots, \textbf{r}_{N}; L).
		\label{homogenius}
	\end{equation}
	Note that the potential energy expressed via the Coulomb potential as well as the Ewald one and AAEP satisfies condition (\ref{homogenius}).
	
	The derivative in Eq.~\eqref{eq:pvkt} now reads:
	\begin{multline}
		\gamma\left(\frac{\partial(\beta U(\gamma\textbf{r}_{1},\ldots, \gamma\textbf{r}_{N}; \gamma L))}{\partial \gamma}\right)_T = {} \\
		\gamma\frac{\partial}{\partial\gamma}\left(
		\frac{1}{\gamma}\beta U(\textbf{r}_{1},\ldots, \textbf{r}_{N}; L)
		\right)_T = {} \\ -\beta U(\gamma\textbf{r}_{1},\ldots, \gamma\textbf{r}_{N}; \gamma L).
	\end{multline}
	Then expression~\eqref{eq:pvkt} is simplified to:
	\begin{equation}
		\label{1+U/3N}
		\frac{\beta P_F V}{N} = 1 +  \frac{\left\langle\beta U\right\rangle}{3N} .
	\end{equation}
	
	Thus, we have shown that the total pressure in the Coulomb system of $N$ particles is expressed by Eq.~\eqref{eq:1+U/3N_coulomb}.
	
	Pressure in atomistic simulations is often calculated from Eq.~\eqref{pv:virial}. As it was shown in Sec.~\ref{sec:PBC}, Eq.~\eqref{correct pv:virial} should be used for potentials with an explicit dependence on the cell size instead of Eq.~\eqref{pv:virial}. Below we demonstrate that Eq.~\eqref{correct pv:virial} reduces to Eq.~\eqref{1+U/3N} for the Ewald potential and AAEP.

	The following expression should be satisfied for formulas~\eqref{1+U/3N} and~\eqref{correct pv:virial} to coincide:
	\begin{multline}
		\label{eq:criteria}
		f[U] \coloneqq \frac{1}{2} \sum_{i=1}^{N} \sum_{\substack{j=1 \\ j\neq i}}^{N} \textbf{r}_{ij} \cdot \left(\frac{\partial U}{\partial \textbf{r}_{ij}}\right)_{T, L} 
		+ L\left(\frac{\partial U}{\partial L}\right)_{T, \textbf{r}_i} \\ {} = -U.
	\end{multline}
	The functional $f[U]$ is linear in $U$: $f[U_1 + U_2] = f[U_1]~+~f[U_2]$, so the fulfillment of criterion~\eqref{eq:criteria} can be demonstrated individually for $U_0$ and $U_\text{pair}$. 
	
	\subsection{Pressure for the Ewald potential}
	\label{sec:ewald}

	In this subsection, we consider the Ewald potential (see Eqs. \eqref{v_1}, \eqref{v_2}). Let us compute the first term of $f[U_\text{pair}]$ (see Eq. \eqref{eq:criteria}):
	\begin{equation}
		\label{eq:ew_vir}
		\frac{1}{2} \sum_{i=1}^{N} \sum_{\substack{j=1 \\ j\neq i}}^{N}
		\textbf{r}_{ij} \cdot \left(\frac{\partial U_\text{pair}}{\partial \textbf{r}_{ij}}\right)_{T, L} = 
		-
		\frac{1}{2}\sum_{i=1}^{N} \sum_{\substack{j=1 \\ j\neq i}}^{N}
		q_i q_j F\left( \textbf{r}_{ij} \right),
	\end{equation}
	and the second one
	\begin{equation}
		\label{eq:ew_der}
		L\left(\frac{\partial U_\text{pair}}{\partial L}\right)_{T, \textbf{r}_i} = -U_\text{pair} + \frac{1}{2}\sum_{i=1}^{N} \sum_{\substack{j=1 \\ j\neq i}}^{N}q_i q_j F\left( \textbf{r}_{ij} \right).
	\end{equation}
	Here,
	\begin{multline}
		F\left( \textbf{r} \right) = \frac{1}{L^2}\sum_{\textbf{n}} \frac{G\left(\left|\textbf{r}/L + \textbf{n} \right|\right)}{\left|\textbf{r}/L + \textbf{n} \right|} \left( r^2/L + \textbf{r} \cdot \textbf{n} \right) \\
		+ \sum_{\textbf{n} \neq \textbf{0}} \frac{e^{-\pi^2 n^2/\delta^2}}{\pi n^2} \frac{2\pi}{L^2}(\textbf{r} \cdot \textbf{n})\sin \left( \frac{2\pi}{L} \textbf{n} \cdot \textbf{r} \right),
	\end{multline}
	and (see Eq. (4.9) in Ref.~\cite{Liang_2022}):
	\begin{equation}
		\label{G(x)}
		G\left( x \right) =  \frac{2\delta e^{-\delta^2 x^2} }{\sqrt{\pi}x } + \frac{\erfc\left(\delta x\right)}{x^2} .
	\end{equation}
	Thus, $f[U_\text{pair}] =~\eqref{eq:ew_vir} +~\eqref{eq:ew_der} = - U_\text{pair}$. 
	
	Now we show that criterion~\eqref{eq:criteria} for constant contributions in TCP and OCP is satisfied:
	\begin{equation}
		f[U_0^{\mathrm{TCP}}] = L \frac{\partial }{\partial L}\left( \frac{\chi}{2L} \sum_{i=1}^N q_i^2 \right)_{T, \textbf{r}_i} = - U_0^{\mathrm{TCP}},
	\end{equation}
	\begin{equation}
		f[U_0^{\mathrm{OCP}}] = L \frac{\partial }{\partial L}\left( \frac{\chi- \pi N/\delta^2}{2L} (Ze)^2N \right)_{T, \textbf{r}_i} = - U_0^{\mathrm{OCP}}.
	\end{equation}
	Therefore, criterion~\eqref{eq:criteria} is also satisfied for the constant contributions and $f[U] = f[U_0 + U_\text{pair}] =  -U_0 - U_\text{pair} = -U$; thus, formulas~\eqref{1+U/3N} and~\eqref{correct pv:virial} coincide. Consequently, the OCCS and TCCS pressure is expressed via the potential energy, $U/3$.

	In a recent paper~\cite{Liang_2022}, the pressure of a Coulomb system using the Ewald potential was also considered. However, Eq.~(4.13) in Ref.~\cite{Liang_2022} for the pressure fails to reduce to the simple form~\eqref{1+U/3N}. Although the dependence of the potential on the cell length is taken into account in~\cite{Liang_2022}, Eq.~(4.13) in~\cite{Liang_2022} turned out to be incorrect. 
	
	Expression (4.8) in Ref.~\cite{Liang_2022} for the potential gradient is correct and was taken from a previously published article~\cite{Correct_grad}.
	However, Eqs.  (4.10), (4.11) in Ref.~\cite{Liang_2022} for the partial derivative of potential energy with respect to  volume are wrong. Next, we explain and eliminate errors made in the derivation of Eq.~(4.13) in Ref.~\cite{Liang_2022}.
	
	First, during the Ewald summation in Ref.~\cite{Liang_2022}, the following expression is used (see Eq. (4.3) in Ref.~\cite{Liang_2022}):
	\begin{equation}
		\label{eq:CoulombErfErfc}
		\cfrac{1}{r} = \cfrac{\erf(\sqrt{\alpha}r)}{r} + \cfrac{\erfc(\sqrt{\alpha}r)}{r},
	\end{equation}
	where ``$\alpha$ is a positive constant''. Although Eq. \eqref{eq:CoulombErfErfc} is correct, the quantity $\sqrt{\alpha}$ has a \emph{length dimension}. Since $r \propto L$, $\alpha \propto 1/L^2$. To connect Eqs. (4.5), (4.6) in~\cite{Liang_2022} with Eqs. \eqref{v_1}, \eqref{v_2}, we should write $\alpha = \delta^2 / L^2$. Thus, the derivative of $\alpha$ over  volume should be taken into account, which is not done during the derivation of Eqs. (4.10)--(4.13) in Ref.~\cite{Liang_2022}.
	
	Second, while differentiating the potential energy over volume in Eqs. (4.10) and (4.11) of Ref.~\cite{Liang_2022}, the authors fix the dimensionless coordinates $\textbf{s}_i = \textbf{r}_i/L$, whereas the usual coordinates $\textbf{r}_i$ should be fixed. Thus, instead of calculating $\left(\partial U/\partial L\right)_{T, \textbf{r}_i}$ in Eq. \eqref{eq:dU/dV} (or in Eq. (3.8) in Ref.~\cite{Liang_2022}), the authors of~\cite{Liang_2022} evaluate $\left(\partial U/\partial L\right)_{T, \textbf{s}_i}$. This error leads to an incorrect pressure contribution associated with the explicit dependence of the potential energy on volume.

	
	Next, we derive the correct volume derivatives in the notations of Ref.~\cite{Liang_2022}.
	
	The potential energy of TCP (see Eqs. (4.5), (4.6) in Ref.~\cite{Liang_2022}) is expressed in terms of the complex quantity $\rho(\textbf{k})$:
	\begin{equation}
		\rho(\textbf{k}) = \sum_{i=1}^N q_i \exp(i \textbf{k} \cdot \textbf{r}_i).
	\end{equation}
	Expanding the modulus $|\rho(\textbf{k})|^2$ in Eq. (4.6) of Ref.~\cite{Liang_2022}, one can obtain the potential energy in the form of Eq.~\eqref{eq:TCP_full} (see also Eqs. (1.4)--(1.7) in Ref.~\cite{DeLeeuw1980SimulationOE}).
	
	Next, we rewrite the formulas for the TCP potential energy, $U = U_1 + U_2$, given in Ref.~\cite{Liang_2022} (see Eqs.~(4.5), (4.6)):
	\begin{equation}
		\label{U_1}
		U_1 = \frac{2\pi}{V} \sum_{\textbf{k} \neq \textbf{0}}\frac{e^{-\frac{k^2}{4 \alpha}}}{k^2}\left|\rho(\textbf{k})\right|^2 - \sum_{i=1}^N q_i^2 \sqrt{\frac{\alpha}{\pi}},  
	\end{equation}
	where the summation is performed over $\textbf{k} = 2\pi \textbf{n}/L$, and
	\begin{equation}
		\label{U_2}
		U_2 = \frac{1}{2}\sum_{i=1}^{N}\sum_{j=1}^{N}q_i q_j \sideset{}{'}\sum_{\textbf{n}} \frac{\erfc\left( \alpha \left|\textbf{r}_{ij} + \textbf{n}L \right|\right)}{\left| \textbf{r}_{ij} + \textbf{n}L\right|}.
	\end{equation}
	Now we calculate the partial volume derivatives of the expressions~\eqref{U_1},~\eqref{U_2} with fixed particle positions:
	\begin{multline}
		\label{deru1}
		\left(\frac{\partial U_1}{\partial V}\right)_{T, \textbf{r}_i}
		=
		\frac{1}{3L^2}\left(\frac{\partial U_1}{\partial L}\right)_{T, \textbf{r}_i}
		=
		-\frac{U_1}{3V}\\  + \frac{2\pi}{3V} \sum_{\textbf{k} \neq \textbf{0}}\frac{e^{-\frac{k^2}{4 \alpha}}}{k^2}  \sum_{i=1}^N\sum_{j=1}^Nq_i q_j (\textbf{r}_{ij}\cdot\textbf{k})
		\sin(\textbf{k}\cdot\textbf{r}_{ij})
		,
	\end{multline}
	
	\begin{multline}
		\label{deru2}
		\left(\frac{\partial U_2}{\partial V}\right)_{T, \textbf{r}_i} =
		\frac{1}{3L^2}\left(\frac{\partial U_2}{\partial L}\right)_{T, \textbf{r}_i} =
		-\frac{U_2}{3V} + \frac{1}{6V}   \\ \times \sum_{i=1}^{N}\sum_{j=1}^{N} \frac{q_i q_j}{L} \sideset{}{'}\sum_{\textbf{n}} \left(\frac{\textbf{n}\cdot \textbf{r}_{ij}}{L} + \frac{r^2_{ij}}{L^2}\right)\frac{G\left(\left|\textbf{r}_{ij}/L + \textbf{n} \right|\right)}{\left|\textbf{r}_{ij}/L + \textbf{n} \right|}.
	\end{multline}
	A detailed derivation of Eqs. \eqref{deru1} and \eqref{deru2} can be found in App.~\ref{app:LiangDeriv}.

		The forces $\textbf{f}_i = \textbf{f}_{i,1} + \textbf{f}_{i,2} $ acting on the $i$th particle was correctly obtained in Ref.~\cite{Liang_2022} (see Eq.(4.8)):
	
	\begin{multline}
		\textbf{f}_{i, 1}  = -\frac{4\pi}{V} \sum_{\textbf{k} \neq \textbf{0}} \frac{e^{-\frac{k^2}{4\alpha^2}}}{k^2} q_i \textbf{k} \mathrm{Im}\left( e^{-i\textbf{k} \textbf{r}_i} \rho(\textbf{k})\right) \\ 
		= -\frac{4\pi}{V} \sum_{\textbf{k} \neq \textbf{0}}\frac{e^{-\frac{k^2}{4\alpha^2}}}{k^2} q_i \textbf{k} \sum_{j=1}^N \cos(\textbf{k}\cdot \textbf{r}_{ij}),
	\end{multline}
	
	\begin{equation}
		\textbf{f}_{i,2} =  -\frac{q_i}{L^2} \sideset{}{'}\sum_{j, \textbf{n}} q_j \frac{G(|\textbf{r}_{ij}/L + \textbf{n}|)}{|\textbf{r}_{ij}/L + \textbf{n}|}(\textbf{r}_{ij}/L + \textbf{n}).
	\end{equation}

Thus, if we consider the correct volume derivatives $\left(\partial U/\partial L\right)_{T, \textbf{r}_i}$  calculated above in Eqs.~\eqref{deru1}--\eqref{deru2}, instead of $\left(\partial U/\partial L\right)_{T, \textbf{s}_i}$, and the gradient of $U_1$ and $U_2$ in Eq.~(4.8) of Ref.~\cite{Liang_2022}, formula~(4.13) in Ref.~\cite{Liang_2022}, as expected, is simplified to Eq.~\eqref{1+U/3N}:
\begin{multline}
	\frac{\beta P V}{N} = 1 + \frac{\beta}{3N} \\ \times 
	\left\langle  \sum_{i=1}^N \textbf{r}_i \cdot (\textbf{f}_{i,1}+\textbf{f}_{i,2}) - 
	 L\left(\frac{\partial ( U_1 + U_2)}{\partial L}\right)_{T, \textbf{r}_i} \right\rangle \\
	 = 1 + \frac{\left\langle \beta U \right\rangle}{3N}.
\end{multline}

	\subsection{Pressure for the angular-averaged Ewald potential}
	In this subsection, we show that the formula for the virial pressure~\eqref{correct pv:virial} and Eq.~\eqref{1+U/3N} coincide in the case of the AAEP for OCP and TCP.
	
	To do this, we calculate $f[U^\text{AAEP}_\text{pair}]$, $f[U_{0, a}]$ similar to Sec.~\eqref{sec:ewald}:
	\begin{equation}
		\label{eq:pressureAAEPfirst}
		\frac{1}{2} \sum_{i=1}^{N} \sum_{\substack{j=1 \\ j\neq i}}^{N} 
		\textbf{r}_{ij}\cdot \left(\frac{\partial U_\text{pair}^\text{AAEP}}{\partial \textbf{r}_{ij}}\right)_{T, L}
		\!\!\! =\!
		\sum_{i=1}^{N} \sum_{\substack{j=1 \\ j\neq i}}^{N}
		\frac{q_i q_j}{2} \left( \frac{r^2_{ij}}{r^3_m} -\frac{1}{r_{ij}} \right) ,
	\end{equation}
	\begin{equation}
		\label{eq:pressureAAEPsecond}
		L\left(\frac{\partial U_\text{pair}^\text{AAEP}}{\partial L}\right)_{T, \textbf{r}_i} = \frac{3}{4}\sum_{i=1}^{N} \sum_{\substack{j=1 \\ j\neq i}}^{N}q_i q_j \left( \frac{1}{r_m} - \frac{r^2_{ij}}{r_m^3} \right).
	\end{equation}
	Hence, $f[U_\text{pair}^\text{AAEP}] = \eqref{eq:pressureAAEPfirst}+\eqref{eq:pressureAAEPsecond} = -U_\text{pair}^\text{AAEP}$.
	The constant contributions of the AAEP in the case of TCP and OCP do not depend on the coordinates and are proportional to $L^{-1}$. Therefore
	\begin{equation}
		f[U_{0,a}^\text{TCP}] = L\frac{\partial}{\partial L}\left( -\frac{3}{4} \sum_{i=1}^N \frac{q_i^2}{r_m}\right)_{T, \textbf{r}_i} =-U_{0,a}^\text{TCP}, 
	\end{equation}
	
	\begin{equation}
		f[U_{0,a}^\text{OCP}] = L\frac{\partial}{\partial L}\left(-\frac{3}{20} \frac{(Ze)^2 N(N+5)}{L}\right)_{T, \textbf{r}_i} = -U_{0,a}^\text{OCP},
	\end{equation}
	and Eqs.~\eqref{1+U/3N}, \eqref{correct pv:virial} coincide in the case of TCP and OCP interacting via the AAEP.

	\section{Conclusion}
	\label{sec:conc}
	
	In this work, we want to emphasize two main results. First, the expression for the virial pressure should be corrected for potentials with an explicit dependence on the system volume. An important example of such a potential is the Ewald one or its angular-averaged version, AAEP.
	A similar correction should be used in average atom models, where the self-consistent potential depends on the spherical cell volume \cite{Ovechkin:HEDP:2014}.
	
	Second, the Ewald potential and the AAEP are homogeneous functions of particle coordinates and a cell length.  Therefore, the derivative of free energy with respect to volume can be easily calculated. Thus, we obtain the same formula for pressure as for the Coulomb potential: pressure is determined by potential energy. Using the corrected expression for virial pressure, we obtain exactly the same formula. 
	In addition, we explain why the complicated formula (4.13) in Ref.~\cite{Liang_2022} is incorrect and expression \eqref{1+U/3N} should be used instead. 

	Finally, we confirm the statement known since the XIXth century that the pressure of a classical Coulomb system is completely defined by its potential energy.
	
	\begin{acknowledgments}
		The work was supported by the Foundation for the Advancement of Theoretical Physics and Mathematics ``BASIS''.
	\end{acknowledgments}
	
	\appendix
	\section{Derivation of the virial in a pair representation}
	\label{app:virialPair}
	We represent the force acting on the $i$th particle, $\textbf{f}_i$, as the sum of the forces acting on the $i$th particle from all other particles in the cubic cell:
	\begin{equation}
		\textbf{f}_i=\sum_{\substack{j=1 \\ j\neq i}}^N\textbf{f}_{ij},
	\end{equation}
	where $\textbf{f}_{ij}$ is the force acting on the $i$th particle from the $j$th particle. Then we divide the sum in the virial \eqref{virial definition} into two identical sums:
	\begin{equation}
		\label{eq:virialTwoSumhalf}
		W = \sum_{i=1}^{N} \textbf{r}_i \cdot \sum_{\substack{j=1 \\ j\neq i}}^N\textbf{f}_{ij}
		=\frac{1}{2}\sum_{i=1}^{N} \textbf{r}_i \cdot \sum_{\substack{j=1 \\ j\neq i}}^N\textbf{f}_{ij}
		+
		\frac{1}{2}\sum_{i=1}^{N} \textbf{r}_i \cdot \sum_{\substack{j=1 \\ j\neq i}}^N\textbf{f}_{ij},
	\end{equation}
	Next we replace $\textbf{f}_{ij} = - \textbf{f}_{ji}$ (since Newton's third law holds) and change the summation indices in the second term of Eq. \eqref{eq:virialTwoSumhalf}:
	\begin{equation}
		\frac{1}{2}\sum_{i=1}^{N} \textbf{r}_i \cdot \sum_{\substack{j=1 \\ j\neq i}}^N\textbf{f}_{ij}
		=
		-\frac{1}{2}\sum_{i=1}^{N} \textbf{r}_i \cdot \sum_{\substack{j=1 \\ j\neq i}}^N\textbf{f}_{ji}
		=
		-\frac{1}{2}\sum_{i=1}^{N} \textbf{r}_j \cdot \sum_{\substack{j=1 \\ j\neq i}}^N\textbf{f}_{ij}.
	\end{equation}
	Finally, the virial can be represented as
	\begin{equation}
		W = \frac{1}{2}\sum_{i=1}^N\sum_{\substack{j=1\\ j\neq i}}^N \textbf{r}_{i} \cdot\textbf{f}_{ij}   - \frac{1}{2}\sum_{i=1}^N \sum_{\substack{j=1\\ j\neq i}}^N \textbf{r}_{j}\cdot \textbf{f}_{ij} = \frac{1}{2} \sum_{i=1}^N \sum_{\substack{j=1\\ j\neq i}}^N \textbf{r}_{ij} \cdot\textbf{f}_{ij}.
	\end{equation}

	\section{Derivation of OCP energy expression}
	\label{app:OCPenergyDeltaDeriv}
	The energy of OCP in the form of Eq. \eqref{eq:TCP_full} can be obtained from the energy of the Yukawa one-component plasma (YOCP) as the Debye length, $k_D$, tends to zero. We use an expression for the potential energy of YOCP derived in Ref.~\cite{Hamaguchi:JChemPhys:1994} (see Eqs. (24) and (26) in~\cite{Hamaguchi:JChemPhys:1994}):
	\begin{equation}
		U^\text{YOCP} = \frac{1}{2} \sum_{i=1}^N \sum_{\substack{j=1\\ j \neq i}}^N (Ze)^2 \psi(\textbf{r}_{ij}; L) + U^\text{YOCP}_0,
	\end{equation}
	where 
	\begin{equation}
		\label{U0 YOCP}
		U^\text{YOCP}_0 = \frac{1}{2} N (Ze)^2\lim\limits_{r \to 0} \left[\psi(\textbf{r}; L)  - \frac{1}{r}\right]. 
	\end{equation}
	Note that we use Gaussian units.	
	For convenience, we also use the dimensionless screening parameter $\xi$, instead of $k_D$:
	\begin{equation}
		\xi = k_D L.
	\end{equation}

	The Ewald potential for the YOCP, $\psi(\textbf{r}; L)~=\psi_1(\textbf{r}; L)~+~\psi_2(\textbf{r}; L)$, can be written as follows (see Eqs.~(36) and (38) in Ref.~\cite{Hamaguchi:JChemPhys:1994}):	
	\begin{equation}
		\label{psi_1}
		\psi_1(\textbf{r}; L) =  \frac{1}{L} \sum_{\textbf{n}}\frac{\eta\left(\left|\textbf{r}/L + \textbf{n} \right| \right)}{\left|\textbf{r}/L + \textbf{n} \right|}e^{-\xi \left|\textbf{r}/L + \textbf{n} \right| } + C_0,
	\end{equation}
	\begin{multline}
		\label{psi_2}
		\psi_2(\textbf{r}; L) = \frac{1}{L} \frac{e^{-\xi^2/(4\delta^2)}/\sqrt{\pi}}{1 + \erf(\xi/(2\delta))} \sum_{\textbf{n} \neq \textbf{0}} \frac{\cos\left(2 \pi \textbf{n} \cdot \textbf{r}/L \right)}{\xi^2/4 + \pi^2 n^2}\\
		\times \left[\pi^{3/2} e^{-\pi^2 n^2/\delta^2} + \frac{\xi}{n}D\left( \pi n/\delta\right) \right],
	\end{multline}
	where the following notations are used:
	\begin{equation}
		C_0 = 	- \frac{4 \pi}{\xi^2 L} \left[ 1 - \frac{e^{-\xi^2/(4\delta^2)}}{1 + \erf(\xi/(2\delta))}\left(1 + \frac{\xi}{\delta\sqrt{\pi}} \right)  \right],
	\end{equation}
	\begin{equation}
		\eta(x) = \frac{\erfc(\delta x - \xi/(2\delta))}{1 + \erf(\xi/(2\delta))},
	\end{equation}
	and $D(x)$ is the Dawson function:
	\begin{equation}
		D(x) = e^{-x^2}\int_0^x e^{t^2}dt.
	\end{equation}
	
	As the screening parameter tends to zero, $\xi \to 0$, one can obtain the Ewald potential for the OCP from the one for the YOCP:
	\begin{equation}
		\lim\limits_{\xi \to 0} C_0
		= -\frac{\pi}{L \delta^2},
	\end{equation}
	\begin{equation}
		\lim\limits_{\xi \to 0} \psi_1(\textbf{r}; L) = 
		v_1(\textbf{r}; L) - \frac{\pi}{L \delta^2},
	\end{equation}
	
	\begin{equation}
		\lim\limits_{\xi \to 0} \psi_2(\textbf{r}; L) = v_2(\textbf{r}; L).
	\end{equation}
	Using the above limits, we obtain the constant contribution:
	\begin{equation}
		\label{u01}
		\lim\limits_{\xi \to 0} U^\text{YOCP}_0 = \frac{1}{2}N(Ze)^2 \frac{\chi - \pi/\delta^2}{L}=\tilde{U}^\text{OCP}_0.
	\end{equation}
	
	Thus, the potential energy of OCP is the following:
	\begin{multline}
		U^\text{OCP}= \frac{1}{2} \sum_{i=1}^N \sum_{\substack{j=1\\ j \neq i}}^N (Ze)^2 \left[v(\textbf{r};L)-\cfrac{\pi}{L\delta^2}\right]
		+\tilde{U}^\text{OCP}_0
		\\=
		\frac{1}{2} \sum_{i=1}^N \sum_{\substack{j=1\\ j \neq i}}^N (Ze)^2 v(\textbf{r};L)+U_0^{\mathrm{OCP}},
	\end{multline}
	where
	\begin{multline}
		\label{eq:U0_OCP_deriv}
		U_0^{\mathrm{OCP}} = \frac{1}{2}N(Ze)^2 \frac{\chi - \pi/\delta^2}{L} - 
		\frac{(Ze)^2}{2}\cfrac{\pi}{L\delta^2} \sum_{i=1}^N \sum_{\substack{j=1\\ j \neq i}}^N 1 \\ =
		\frac{1}{2}N(Ze)^2 \frac{\chi - \pi/\delta^2}{L}
		-
		\frac{(Ze)^2}{2}\cfrac{\pi}{L\delta^2}N(N - 1) \\= \frac{\chi - \pi N/\delta^2}{2L} (Ze)^2N.
	\end{multline}
	Therefore, expression \eqref{eq:U0_OCP_deriv} coincides with Eq.~\eqref{eq:OCP_Ewald_const}.
	
	The Ewald OCP potential and energy with the $\delta$ parameter was also obtained in Ref.~\cite{Fraser:PRB:1996} (see formula (16)) by shifting the Ewald potential, $v(\textbf{r}; L)$, so that its average value over the simulation cell was zero.

	\section{Derivation of derivatives for $U_1$ and $U_2$ in Ref.~\cite{Liang_2022} }
	\label{app:LiangDeriv}
	To differentiate $U_1$ and $U_2$ with respect to the volume, we use the relations  $V = L^3$, $\sqrt{\alpha} = \delta / L$, $\textbf{k} = 2\pi \textbf{n}/L$  and take the derivative by $L$. Finally, the derivative over $L$ should be multiplied by $dL/dV = 1/(3L^2)$. 
	
	First, we calculate the derivative of $|\rho(\textbf{k})|^2$:	
	\begin{multline}
		\label{eq:rhoKDeriv}
		\left(\frac{\partial |\rho(\textbf{k})|^2}{\partial L}\right)_{T, \textbf{r}_i}
		\\=
		\left(\frac{\partial \rho(\textbf{k})}{\partial L}\right)_{T, \textbf{r}_i}\rho^*(\textbf{k}) + \left(\frac{\partial \rho^*(\textbf{k})}{\partial L}\right)_{T, \textbf{r}_i}\rho(\textbf{k}) \\= -\sum_{i=1}^N\sum_{j=1}^N  \frac{i}{L}q_iq_j \textbf{k} \cdot\textbf{r}_i e^{i \textbf{k}\cdot \textbf{r}_{ij}} + \sum_{i=1}^N\sum_{j=1}^N  \frac{i}{L}q_iq_j \textbf{k}\cdot \textbf{r}_i e^{-i \textbf{k}\cdot \textbf{r}_{ij}}
		\\=
		\cfrac{2}{L}\sum_{i=1}^N \textbf{r}_i\cdot\textbf{k}\sum_{j=1}^N  q_i q_j 
		\sin(\textbf{k}\cdot\textbf{r}_{ij})
		\\=
		\cfrac{1}{L}\sum_{i=1}^N\sum_{j=1}^Nq_i q_j (\textbf{r}_{ij}\cdot\textbf{k})
		\sin(\textbf{k}\cdot\textbf{r}_{ij})
		.
	\end{multline}
	To obtain the final equality in Eq. \eqref{eq:rhoKDeriv}, we used the same trick as in App. \ref{app:virialPair}.
	
	Therefore, the derivative of $U_1$ with respect to $L$ equals to:
	\begin{multline}
		\label{eq:deru2L}
		\left(\frac{\partial U_1}{\partial L}\right)_{T, \textbf{r}_i} \\= \frac{\partial}{\partial L}\Bigg(\frac{2\pi}{L} \sum_{\textbf{n} \neq \textbf{0}}\frac{e^{-\frac{\pi^2 n^2}{\delta^2}}}{4 \pi^2 n^2}\left|\rho\left(\frac{2\pi\textbf{n}}{L}\right)\right|^2 - \sum_{i=1}^N q_i^2 \sqrt{\frac{\delta^2}{\pi L^2}} \Bigg) \\
		= \frac{1}{L}\sum_{i=1}^N q_i^2 \sqrt{\frac{\alpha}{\pi}} - \frac{2\pi}{L^2} \sum_{\textbf{k} \neq \textbf{0}}\frac{e^{-\frac{k^2}{4\alpha}}}{k^2}|\rho(\textbf{k})|^2 \\+\frac{2\pi}{L^3} \sum_{\textbf{k} \neq \textbf{0}}\frac{e^{-\frac{k^2}{4\alpha}}}{k^2}\left(\frac{\partial |\rho(\textbf{k})|^2}{\partial L}\right)_{T, \textbf{r}_i} 
		\\=
		-\frac{U_1}{L}  + \frac{2\pi}{L^4} \sum_{\textbf{k} \neq \textbf{0}}\frac{e^{-\frac{k^2}{4 \alpha}}}{k^2}  \sum_{i=1}^N\sum_{j=1}^Nq_i q_j (\textbf{r}_{ij}\cdot\textbf{k})
		\sin(\textbf{k}\cdot\textbf{r}_{ij})
		.
	\end{multline}
	
	Next, we calculate the derivative of each term in the sum \eqref{U_2} with respect to $L$:
	\begin{multline}
		\frac{\partial }{\partial L}\left( \frac{\erfc\left( \alpha \left|\textbf{r} + \textbf{n}L \right|\right)}{\left| \textbf{r} + \textbf{n}L\right|} \right) \equiv\frac{\partial }{\partial L} \left( \frac{1}{L} \frac{\erfc\left( \delta \left|\textbf{r}/L + \textbf{n} \right|\right)}{\left| \textbf{r}/L + \textbf{n}\right|}  \right) \\ = 
		-\frac{1}{L^2} \frac{\erfc\left( \delta \left|\textbf{r}/L + \textbf{n} \right|\right)}{\left| \textbf{r}/L + \textbf{n}\right|} 
		\\+ \frac{1}{L^3}\left(\textbf{n}\cdot \textbf{r} + r^2/L\right)\frac{G\left(\left|\textbf{r}/L + \textbf{n} \right|\right)}{\left|\textbf{r}/L + \textbf{n} \right|},
	\end{multline}
	to obtain the derivative of $U_2$ with respect to $L$:
	\begin{multline}
		\label{eq:deru1L}
		\left(\frac{\partial U_2}{\partial L}\right)_{T, \textbf{r}_i} = -\frac{U_2}{L} \\+ \frac{1}{2L^2}\sum_{i=1}^{N}\sum_{j=1}^{N} q_i q_j \sideset{}{'}\sum_{\textbf{n}} (\textbf{n}\cdot \textbf{r}_{ij}/L + r_{ij}^2/L^2)\frac{G\left(\left|\textbf{r}_{ij}/L + \textbf{n} \right|\right)}{\left|\textbf{r}_{ij}/L + \textbf{n} \right|}.
	\end{multline}
	
	Multiplying \eqref{eq:deru1L} and \eqref{eq:deru2L} by $1/(3L^2)$ one can obtain the derivatives by volume \eqref{deru1} and \eqref{deru2}.
	

\begin{thebibliography}{42}%
\makeatletter
\providecommand \@ifxundefined [1]{%
 \@ifx{#1\undefined}
}%
\providecommand \@ifnum [1]{%
 \ifnum #1\expandafter \@firstoftwo
 \else \expandafter \@secondoftwo
 \fi
}%
\providecommand \@ifx [1]{%
 \ifx #1\expandafter \@firstoftwo
 \else \expandafter \@secondoftwo
 \fi
}%
\providecommand \natexlab [1]{#1}%
\providecommand \enquote  [1]{``#1''}%
\providecommand \bibnamefont  [1]{#1}%
\providecommand \bibfnamefont [1]{#1}%
\providecommand \citenamefont [1]{#1}%
\providecommand \href@noop [0]{\@secondoftwo}%
\providecommand \href [0]{\begingroup \@sanitize@url \@href}%
\providecommand \@href[1]{\@@startlink{#1}\@@href}%
\providecommand \@@href[1]{\endgroup#1\@@endlink}%
\providecommand \@sanitize@url [0]{\catcode `\\12\catcode `\$12\catcode
  `\&12\catcode `\#12\catcode `\^12\catcode `\_12\catcode `\%12\relax}%
\providecommand \@@startlink[1]{}%
\providecommand \@@endlink[0]{}%
\providecommand \url  [0]{\begingroup\@sanitize@url \@url }%
\providecommand \@url [1]{\endgroup\@href {#1}{\urlprefix }}%
\providecommand \urlprefix  [0]{URL }%
\providecommand \Eprint [0]{\href }%
\providecommand \doibase [0]{https://doi.org/}%
\providecommand \selectlanguage [0]{\@gobble}%
\providecommand \bibinfo  [0]{\@secondoftwo}%
\providecommand \bibfield  [0]{\@secondoftwo}%
\providecommand \translation [1]{[#1]}%
\providecommand \BibitemOpen [0]{}%
\providecommand \bibitemStop [0]{}%
\providecommand \bibitemNoStop [0]{.\EOS\space}%
\providecommand \EOS [0]{\spacefactor3000\relax}%
\providecommand \BibitemShut  [1]{\csname bibitem#1\endcsname}%
\let\auto@bib@innerbib\@empty
\bibitem [{\citenamefont {Attard}(2002)}]{Attard:CS:2002}%
  \BibitemOpen
  \bibfield  {author} {\bibinfo {author} {\bibfnamefont {P.}~\bibnamefont
  {Attard}},\ }in\ \href
  {https://doi.org/https://doi.org/10.1016/B978-012066321-7/50012-X} {\emph
  {\bibinfo {booktitle} {Thermodynamics and Statistical Mechanics}}},\ \bibinfo
  {editor} {edited by\ \bibinfo {editor} {\bibfnamefont {P.}~\bibnamefont
  {Attard}}}\ (\bibinfo  {publisher} {Academic Press},\ \bibinfo {address}
  {London},\ \bibinfo {year} {2002})\ pp.\ \bibinfo {pages}
  {317--350}\BibitemShut {NoStop}%
\bibitem [{\citenamefont {Roper}\ \emph {et~al.}(2022)\citenamefont {Roper},
  \citenamefont {Harkema}, \citenamefont {Sabharwall}, \citenamefont {Riddle},
  \citenamefont {Chisholm}, \citenamefont {Day},\ and\ \citenamefont
  {Marotta}}]{Roper:MoltenSalt:2022}%
  \BibitemOpen
  \bibfield  {author} {\bibinfo {author} {\bibfnamefont {R.}~\bibnamefont
  {Roper}}, \bibinfo {author} {\bibfnamefont {M.}~\bibnamefont {Harkema}},
  \bibinfo {author} {\bibfnamefont {P.}~\bibnamefont {Sabharwall}}, \bibinfo
  {author} {\bibfnamefont {C.}~\bibnamefont {Riddle}}, \bibinfo {author}
  {\bibfnamefont {B.}~\bibnamefont {Chisholm}}, \bibinfo {author}
  {\bibfnamefont {B.}~\bibnamefont {Day}},\ and\ \bibinfo {author}
  {\bibfnamefont {P.}~\bibnamefont {Marotta}},\ }\href
  {https://doi.org/https://doi.org/10.1016/j.anucene.2021.108924} {\bibfield
  {journal} {\bibinfo  {journal} {Annals of Nuclear Energy}\ }\textbf {\bibinfo
  {volume} {169}},\ \bibinfo {pages} {108924} (\bibinfo {year}
  {2022})}\BibitemShut {NoStop}%
\bibitem [{\citenamefont {Bo}\ \emph {et~al.}(2018)\citenamefont {Bo},
  \citenamefont {Ren}, \citenamefont {Xu}, \citenamefont {Du},\ and\
  \citenamefont {Dou}}]{Bo:APX:2018}%
  \BibitemOpen
  \bibfield  {author} {\bibinfo {author} {\bibfnamefont {G.}~\bibnamefont
  {Bo}}, \bibinfo {author} {\bibfnamefont {L.}~\bibnamefont {Ren}}, \bibinfo
  {author} {\bibfnamefont {X.}~\bibnamefont {Xu}}, \bibinfo {author}
  {\bibfnamefont {Y.}~\bibnamefont {Du}},\ and\ \bibinfo {author}
  {\bibfnamefont {S.}~\bibnamefont {Dou}},\ }\href
  {https://doi.org/10.1080/23746149.2018.1446359} {\bibfield  {journal}
  {\bibinfo  {journal} {Advances in Physics: X}\ }\textbf {\bibinfo {volume}
  {3}},\ \bibinfo {pages} {1446359} (\bibinfo {year} {2018})}\BibitemShut
  {NoStop}%
\bibitem [{\citenamefont {Smirnov}(2023)}]{Smirnov:JAppPhys:2023}%
  \BibitemOpen
  \bibfield  {author} {\bibinfo {author} {\bibfnamefont {N.~A.}\ \bibnamefont
  {Smirnov}},\ }\href {https://doi.org/10.1063/5.0158737} {\bibfield  {journal}
  {\bibinfo  {journal} {Journal of Applied Physics}\ }\textbf {\bibinfo
  {volume} {134}},\ \bibinfo {pages} {025901} (\bibinfo {year}
  {2023})}\BibitemShut {NoStop}%
\bibitem [{\citenamefont {Potekhin}\ \emph {et~al.}(2009)\citenamefont
  {Potekhin}, \citenamefont {Chabrier}, \citenamefont {Chugunov}, \citenamefont
  {DeWitt},\ and\ \citenamefont {Rogers}}]{PotekhinAddendum:PRE:2009}%
  \BibitemOpen
  \bibfield  {author} {\bibinfo {author} {\bibfnamefont {A.~Y.}\ \bibnamefont
  {Potekhin}}, \bibinfo {author} {\bibfnamefont {G.}~\bibnamefont {Chabrier}},
  \bibinfo {author} {\bibfnamefont {A.~I.}\ \bibnamefont {Chugunov}}, \bibinfo
  {author} {\bibfnamefont {H.~E.}\ \bibnamefont {DeWitt}},\ and\ \bibinfo
  {author} {\bibfnamefont {F.~J.}\ \bibnamefont {Rogers}},\ }\href
  {https://doi.org/10.1103/PhysRevE.80.047401} {\bibfield  {journal} {\bibinfo
  {journal} {Phys. Rev. E}\ }\textbf {\bibinfo {volume} {80}},\ \bibinfo
  {pages} {047401} (\bibinfo {year} {2009})}\BibitemShut {NoStop}%
\bibitem [{\citenamefont {{Potekhin, A. Y.}}\ and\ \citenamefont {{Chabrier,
  G.}}(2013)}]{Potekhin:AA:2013}%
  \BibitemOpen
  \bibfield  {author} {\bibinfo {author} {\bibnamefont {{Potekhin, A. Y.}}}\
  and\ \bibinfo {author} {\bibnamefont {{Chabrier, G.}}},\ }\href
  {https://doi.org/10.1051/0004-6361/201220082} {\bibfield  {journal} {\bibinfo
   {journal} {A\&A}\ }\textbf {\bibinfo {volume} {550}},\ \bibinfo {pages}
  {A43} (\bibinfo {year} {2013})}\BibitemShut {NoStop}%
\bibitem [{\citenamefont {Fortov}\ \emph {et~al.}(2006)\citenamefont {Fortov},
  \citenamefont {Iakubov},\ and\ \citenamefont {Khrapak}}]{Fortov:NPP:2006}%
  \BibitemOpen
  \bibfield  {author} {\bibinfo {author} {\bibfnamefont {V.}~\bibnamefont
  {Fortov}}, \bibinfo {author} {\bibfnamefont {I.}~\bibnamefont {Iakubov}},\
  and\ \bibinfo {author} {\bibfnamefont {A.}~\bibnamefont {Khrapak}},\ }\href
  {https://doi.org/10.1093/acprof:oso/9780199299805.001.0001} {\emph {\bibinfo
  {title} {Physics of Strongly Coupled Plasma}}}\ (\bibinfo  {publisher}
  {Oxford University Press},\ \bibinfo {year} {2006})\BibitemShut {NoStop}%
\bibitem [{\citenamefont {Redmer}\ and\ \citenamefont
  {Röpke}(2010)}]{Redmer:CPP:2010}%
  \BibitemOpen
  \bibfield  {author} {\bibinfo {author} {\bibfnamefont {R.}~\bibnamefont
  {Redmer}}\ and\ \bibinfo {author} {\bibfnamefont {G.}~\bibnamefont
  {Röpke}},\ }\href {https://doi.org/https://doi.org/10.1002/ctpp.201000079}
  {\bibfield  {journal} {\bibinfo  {journal} {Contributions to Plasma Physics}\
  }\textbf {\bibinfo {volume} {50}},\ \bibinfo {pages} {970} (\bibinfo {year}
  {2010})}\BibitemShut {NoStop}%
\bibitem [{\citenamefont {Lieb}(1976)}]{Lieb:RMP:1976}%
  \BibitemOpen
  \bibfield  {author} {\bibinfo {author} {\bibfnamefont {E.~H.}\ \bibnamefont
  {Lieb}},\ }\href {https://doi.org/10.1103/RevModPhys.48.553} {\bibfield
  {journal} {\bibinfo  {journal} {Reviews of Modern Physics}\ }\textbf
  {\bibinfo {volume} {48}},\ \bibinfo {pages} {553–569} (\bibinfo {year}
  {1976})}\BibitemShut {NoStop}%
\bibitem [{\citenamefont {Baus}\ and\ \citenamefont
  {Hansen}(1980)}]{Baus:PR:1980}%
  \BibitemOpen
  \bibfield  {author} {\bibinfo {author} {\bibfnamefont {M.}~\bibnamefont
  {Baus}}\ and\ \bibinfo {author} {\bibfnamefont {J.-P.}\ \bibnamefont
  {Hansen}},\ }\href
  {https://doi.org/https://doi.org/10.1016/0370-1573(80)90022-8} {\bibfield
  {journal} {\bibinfo  {journal} {Physics Reports}\ }\textbf {\bibinfo {volume}
  {59}},\ \bibinfo {pages} {1} (\bibinfo {year} {1980})}\BibitemShut {NoStop}%
\bibitem [{\citenamefont {Militzer}\ \emph {et~al.}(2021)\citenamefont
  {Militzer}, \citenamefont {Gonz\'alez-Cataldo}, \citenamefont {Zhang},
  \citenamefont {Driver},\ and\ \citenamefont {Soubiran}}]{Militzer:PRE:2021}%
  \BibitemOpen
  \bibfield  {author} {\bibinfo {author} {\bibfnamefont {B.}~\bibnamefont
  {Militzer}}, \bibinfo {author} {\bibfnamefont {F.}~\bibnamefont
  {Gonz\'alez-Cataldo}}, \bibinfo {author} {\bibfnamefont {S.}~\bibnamefont
  {Zhang}}, \bibinfo {author} {\bibfnamefont {K.~P.}\ \bibnamefont {Driver}},\
  and\ \bibinfo {author} {\bibfnamefont {F.~m.~c.}\ \bibnamefont {Soubiran}},\
  }\href {https://doi.org/10.1103/PhysRevE.103.013203} {\bibfield  {journal}
  {\bibinfo  {journal} {Phys. Rev. E}\ }\textbf {\bibinfo {volume} {103}},\
  \bibinfo {pages} {013203} (\bibinfo {year} {2021})}\BibitemShut {NoStop}%
\bibitem [{\citenamefont {Moldabekov}\ \emph {et~al.}(2023)\citenamefont
  {Moldabekov}, \citenamefont {Vorberger}, \citenamefont {Lokamani},\ and\
  \citenamefont {Dornheim}}]{Moldabekov:JChemPhys:2023}%
  \BibitemOpen
  \bibfield  {author} {\bibinfo {author} {\bibfnamefont {Z.~A.}\ \bibnamefont
  {Moldabekov}}, \bibinfo {author} {\bibfnamefont {J.}~\bibnamefont
  {Vorberger}}, \bibinfo {author} {\bibfnamefont {M.}~\bibnamefont
  {Lokamani}},\ and\ \bibinfo {author} {\bibfnamefont {T.}~\bibnamefont
  {Dornheim}},\ }\href {https://doi.org/10.1063/5.0152126} {\bibfield
  {journal} {\bibinfo  {journal} {The Journal of Chemical Physics}\ }\textbf
  {\bibinfo {volume} {159}},\ \bibinfo {pages} {014107} (\bibinfo {year}
  {2023})}\BibitemShut {NoStop}%
\bibitem [{\citenamefont {Dornheim}\ \emph {et~al.}(2018)\citenamefont
  {Dornheim}, \citenamefont {Groth},\ and\ \citenamefont
  {Bonitz}}]{Dornheim:PR:2018}%
  \BibitemOpen
  \bibfield  {author} {\bibinfo {author} {\bibfnamefont {T.}~\bibnamefont
  {Dornheim}}, \bibinfo {author} {\bibfnamefont {S.}~\bibnamefont {Groth}},\
  and\ \bibinfo {author} {\bibfnamefont {M.}~\bibnamefont {Bonitz}},\ }\href
  {https://doi.org/https://doi.org/10.1016/j.physrep.2018.04.001} {\bibfield
  {journal} {\bibinfo  {journal} {Physics Reports}\ }\textbf {\bibinfo {volume}
  {744}},\ \bibinfo {pages} {1} (\bibinfo {year} {2018})}\BibitemShut {NoStop}%
\bibitem [{\citenamefont {Demyanov}\ and\ \citenamefont
  {Levashov}(2022{\natexlab{a}})}]{DemyanovOCP:PRE:2022}%
  \BibitemOpen
  \bibfield  {author} {\bibinfo {author} {\bibfnamefont {G.~S.}\ \bibnamefont
  {Demyanov}}\ and\ \bibinfo {author} {\bibfnamefont {P.~R.}\ \bibnamefont
  {Levashov}},\ }\href {https://doi.org/10.1103/PhysRevE.106.015204} {\bibfield
   {journal} {\bibinfo  {journal} {Phys. Rev. E}\ }\textbf {\bibinfo {volume}
  {106}},\ \bibinfo {pages} {015204} (\bibinfo {year}
  {2022}{\natexlab{a}})}\BibitemShut {NoStop}%
\bibitem [{\citenamefont {Allen}\ and\ \citenamefont
  {Tildesley}(1987)}]{AllTID}%
  \BibitemOpen
  \bibfield  {author} {\bibinfo {author} {\bibfnamefont {M.~P.}\ \bibnamefont
  {Allen}}\ and\ \bibinfo {author} {\bibfnamefont {D.~J.}\ \bibnamefont
  {Tildesley}},\ }\href@noop {} {\emph {\bibinfo {title} {{Computer simulation
  of liquids}}}}\ (\bibinfo  {publisher} {Oxford university press},\ \bibinfo
  {year} {1987})\BibitemShut {NoStop}%
\bibitem [{\citenamefont {Frenkel}\ and\ \citenamefont
  {Smit}(2002)}]{Frenkel:MD:2002}%
  \BibitemOpen
  \bibfield  {author} {\bibinfo {author} {\bibfnamefont {D.}~\bibnamefont
  {Frenkel}}\ and\ \bibinfo {author} {\bibfnamefont {B.}~\bibnamefont {Smit}},\
  }\href@noop {} {\emph {\bibinfo {title} {Understanding molecular simulation:
  from algorithms to applications}}}\ (\bibinfo  {publisher} {Academic Press
  San Diego},\ \bibinfo {year} {2002})\BibitemShut {NoStop}%
\bibitem [{\citenamefont {Filinov}\ \emph {et~al.}(2023)\citenamefont
  {Filinov}, \citenamefont {Ara},\ and\ \citenamefont
  {Tkachenko}}]{Filinov:PRB:2023}%
  \BibitemOpen
  \bibfield  {author} {\bibinfo {author} {\bibfnamefont {A.~V.}\ \bibnamefont
  {Filinov}}, \bibinfo {author} {\bibfnamefont {J.}~\bibnamefont {Ara}},\ and\
  \bibinfo {author} {\bibfnamefont {I.~M.}\ \bibnamefont {Tkachenko}},\ }\href
  {https://doi.org/10.1103/PhysRevB.107.195143} {\bibfield  {journal} {\bibinfo
   {journal} {Phys. Rev. B}\ }\textbf {\bibinfo {volume} {107}},\ \bibinfo
  {pages} {195143} (\bibinfo {year} {2023})}\BibitemShut {NoStop}%
\bibitem [{\citenamefont {Rapaport}(2004)}]{Rapaport:MD:2004}%
  \BibitemOpen
  \bibfield  {author} {\bibinfo {author} {\bibfnamefont {D.~C.}\ \bibnamefont
  {Rapaport}},\ }\href@noop {} {\emph {\bibinfo {title} {The art of molecular
  dynamics simulation}}}\ (\bibinfo  {publisher} {Cambridge university press},\
  \bibinfo {year} {2004})\BibitemShut {NoStop}%
\bibitem [{\citenamefont {Ewald}(1921)}]{Ewald:1921}%
  \BibitemOpen
  \bibfield  {author} {\bibinfo {author} {\bibfnamefont {P.~P.}\ \bibnamefont
  {Ewald}},\ }\href {https://doi.org/https://doi.org/10.1002/andp.19213690304}
  {\bibfield  {journal} {\bibinfo  {journal} {Annalen der Physik}\ }\textbf
  {\bibinfo {volume} {369}},\ \bibinfo {pages} {253} (\bibinfo {year}
  {1921})}\BibitemShut {NoStop}%
\bibitem [{\citenamefont {Landau}\ and\ \citenamefont
  {Lifshitz}(2013)}]{LandauLifshits:V5:2013}%
  \BibitemOpen
  \bibfield  {author} {\bibinfo {author} {\bibfnamefont {L.~D.}\ \bibnamefont
  {Landau}}\ and\ \bibinfo {author} {\bibfnamefont {E.~M.}\ \bibnamefont
  {Lifshitz}},\ }\href@noop {} {\emph {\bibinfo {title} {Statistical Physics:
  Volume 5}}},\ Vol.~\bibinfo {volume} {5}\ (\bibinfo  {publisher} {Elsevier},\
  \bibinfo {year} {2013})\BibitemShut {NoStop}%
\bibitem [{\citenamefont {Yakub}\ and\ \citenamefont
  {Ronchi}(2003)}]{Yakub:2003}%
  \BibitemOpen
  \bibfield  {author} {\bibinfo {author} {\bibfnamefont {E.}~\bibnamefont
  {Yakub}}\ and\ \bibinfo {author} {\bibfnamefont {C.}~\bibnamefont {Ronchi}},\
  }\href {https://doi.org/10.1063/1.1624364} {\bibfield  {journal} {\bibinfo
  {journal} {The Journal of Chemical Physics}\ }\textbf {\bibinfo {volume}
  {119}},\ \bibinfo {pages} {11556} (\bibinfo {year} {2003})}\BibitemShut
  {NoStop}%
\bibitem [{\citenamefont {Demyanov}\ and\ \citenamefont
  {Levashov}(2022{\natexlab{b}})}]{Demyanov:JPhysA:2022}%
  \BibitemOpen
  \bibfield  {author} {\bibinfo {author} {\bibfnamefont {G.~S.}\ \bibnamefont
  {Demyanov}}\ and\ \bibinfo {author} {\bibfnamefont {P.~R.}\ \bibnamefont
  {Levashov}},\ }\href {https://doi.org/10.1088/1751-8121/ac870b} {\bibfield
  {journal} {\bibinfo  {journal} {Journal of Physics A: Mathematical and
  Theoretical}\ }\textbf {\bibinfo {volume} {55}},\ \bibinfo {pages} {385202}
  (\bibinfo {year} {2022}{\natexlab{b}})}\BibitemShut {NoStop}%
\bibitem [{\citenamefont {{Louwerse}}\ and\ \citenamefont
  {{Baerends}}(2006)}]{Louwerse2006}%
  \BibitemOpen
  \bibfield  {author} {\bibinfo {author} {\bibfnamefont {M.~J.}\ \bibnamefont
  {{Louwerse}}}\ and\ \bibinfo {author} {\bibfnamefont {E.~J.}\ \bibnamefont
  {{Baerends}}},\ }\href {https://doi.org/10.1016/j.cplett.2006.01.087}
  {\bibfield  {journal} {\bibinfo  {journal} {Chemical Physics Letters}\
  }\textbf {\bibinfo {volume} {421}},\ \bibinfo {pages} {138} (\bibinfo {year}
  {2006})}\BibitemShut {NoStop}%
\bibitem [{\citenamefont {Thompson}\ \emph {et~al.}(2009)\citenamefont
  {Thompson}, \citenamefont {Plimpton},\ and\ \citenamefont
  {Mattson}}]{Thompson:JCP:2009}%
  \BibitemOpen
  \bibfield  {author} {\bibinfo {author} {\bibfnamefont {A.~P.}\ \bibnamefont
  {Thompson}}, \bibinfo {author} {\bibfnamefont {S.~J.}\ \bibnamefont
  {Plimpton}},\ and\ \bibinfo {author} {\bibfnamefont {W.}~\bibnamefont
  {Mattson}},\ }\href {https://doi.org/10.1063/1.3245303} {\bibfield  {journal}
  {\bibinfo  {journal} {The Journal of chemical physics}\ }\textbf {\bibinfo
  {volume} {131}} (\bibinfo {year} {2009})}\BibitemShut {NoStop}%
\bibitem [{\citenamefont {Brush}\ \emph {et~al.}(1966)\citenamefont {Brush},
  \citenamefont {Sahlin},\ and\ \citenamefont {Teller}}]{BST:1966}%
  \BibitemOpen
  \bibfield  {author} {\bibinfo {author} {\bibfnamefont {S.~G.}\ \bibnamefont
  {Brush}}, \bibinfo {author} {\bibfnamefont {H.~L.}\ \bibnamefont {Sahlin}},\
  and\ \bibinfo {author} {\bibfnamefont {E.}~\bibnamefont {Teller}},\ }\href
  {https://doi.org/10.1063/1.1727895} {\bibfield  {journal} {\bibinfo
  {journal} {The Journal of Chemical Physics}\ }\textbf {\bibinfo {volume}
  {45}},\ \bibinfo {pages} {2102} (\bibinfo {year} {1966})}\BibitemShut
  {NoStop}%
\bibitem [{\citenamefont {Nikiforov}\ \emph {et~al.}(2005)\citenamefont
  {Nikiforov}, \citenamefont {Novikov},\ and\ \citenamefont
  {Uvarov}}]{Nikiforov:book:2005}%
  \BibitemOpen
  \bibfield  {author} {\bibinfo {author} {\bibfnamefont {A.~F.}\ \bibnamefont
  {Nikiforov}}, \bibinfo {author} {\bibfnamefont {V.~G.}\ \bibnamefont
  {Novikov}},\ and\ \bibinfo {author} {\bibfnamefont {V.~B.}\ \bibnamefont
  {Uvarov}},\ }\href {https://doi.org/10.1007/b137687} {\emph {\bibinfo {title}
  {Quantum-Statistical Models of Hot Dense Matter: Methods for Computation
  Opacity and Equation of State}}},\ Vol.~\bibinfo {volume} {37}\ (\bibinfo
  {publisher} {Springer Science \& Business Media},\ \bibinfo {year}
  {2005})\BibitemShut {NoStop}%
\bibitem [{\citenamefont {Shemyakin}\ \emph {et~al.}(2010)\citenamefont
  {Shemyakin}, \citenamefont {Levashov}, \citenamefont {Obruchkova},\ and\
  \citenamefont {Khishchenko}}]{Shemyakin:JPA:2010}%
  \BibitemOpen
  \bibfield  {author} {\bibinfo {author} {\bibfnamefont {O.~P.}\ \bibnamefont
  {Shemyakin}}, \bibinfo {author} {\bibfnamefont {P.~R.}\ \bibnamefont
  {Levashov}}, \bibinfo {author} {\bibfnamefont {L.~R.}\ \bibnamefont
  {Obruchkova}},\ and\ \bibinfo {author} {\bibfnamefont {K.~V.}\ \bibnamefont
  {Khishchenko}},\ }\href {https://doi.org/10.1088/1751-8113/43/33/335003}
  {\bibfield  {journal} {\bibinfo  {journal} {Journal of Physics A:
  Mathematical and Theoretical}\ }\textbf {\bibinfo {volume} {43}},\ \bibinfo
  {pages} {335003} (\bibinfo {year} {2010})}\BibitemShut {NoStop}%
\bibitem [{\citenamefont {Shemyakin}\ \emph {et~al.}(2019)\citenamefont
  {Shemyakin}, \citenamefont {Levashov},\ and\ \citenamefont
  {Krasnova}}]{Shemyakin:CPC:2019}%
  \BibitemOpen
  \bibfield  {author} {\bibinfo {author} {\bibfnamefont {O.}~\bibnamefont
  {Shemyakin}}, \bibinfo {author} {\bibfnamefont {P.}~\bibnamefont
  {Levashov}},\ and\ \bibinfo {author} {\bibfnamefont {P.}~\bibnamefont
  {Krasnova}},\ }\href
  {https://doi.org/https://doi.org/10.1016/j.cpc.2018.09.008} {\bibfield
  {journal} {\bibinfo  {journal} {Computer Physics Communications}\ }\textbf
  {\bibinfo {volume} {235}},\ \bibinfo {pages} {378} (\bibinfo {year}
  {2019})}\BibitemShut {NoStop}%
\bibitem [{\citenamefont {Liberman}(1979)}]{liberman:PRB:1979}%
  \BibitemOpen
  \bibfield  {author} {\bibinfo {author} {\bibfnamefont {D.~A.}\ \bibnamefont
  {Liberman}},\ }\href {https://doi.org/10.1103/PhysRevB.20.4981} {\bibfield
  {journal} {\bibinfo  {journal} {Phys. Rev. B}\ }\textbf {\bibinfo {volume}
  {20}},\ \bibinfo {pages} {4981} (\bibinfo {year} {1979})}\BibitemShut
  {NoStop}%
\bibitem [{\citenamefont {Ovechkin}\ \emph {et~al.}(2014)\citenamefont
  {Ovechkin}, \citenamefont {Loboda}, \citenamefont {Novikov}, \citenamefont
  {Grushin},\ and\ \citenamefont {Solomyannaya}}]{Ovechkin:HEDP:2014}%
  \BibitemOpen
  \bibfield  {author} {\bibinfo {author} {\bibfnamefont {A.}~\bibnamefont
  {Ovechkin}}, \bibinfo {author} {\bibfnamefont {P.}~\bibnamefont {Loboda}},
  \bibinfo {author} {\bibfnamefont {V.}~\bibnamefont {Novikov}}, \bibinfo
  {author} {\bibfnamefont {A.}~\bibnamefont {Grushin}},\ and\ \bibinfo {author}
  {\bibfnamefont {A.}~\bibnamefont {Solomyannaya}},\ }\href
  {https://doi.org/https://doi.org/10.1016/j.hedp.2014.09.001} {\bibfield
  {journal} {\bibinfo  {journal} {High Energy Density Physics}\ }\textbf
  {\bibinfo {volume} {13}},\ \bibinfo {pages} {20} (\bibinfo {year}
  {2014})}\BibitemShut {NoStop}%
\bibitem [{\citenamefont {Liang}\ \emph {et~al.}(2022)\citenamefont {Liang},
  \citenamefont {Tan}, \citenamefont {Hong}, \citenamefont {Jin}, \citenamefont
  {Xu},\ and\ \citenamefont {Li}}]{Liang_2022}%
  \BibitemOpen
  \bibfield  {author} {\bibinfo {author} {\bibfnamefont {J.}~\bibnamefont
  {Liang}}, \bibinfo {author} {\bibfnamefont {P.}~\bibnamefont {Tan}}, \bibinfo
  {author} {\bibfnamefont {L.}~\bibnamefont {Hong}}, \bibinfo {author}
  {\bibfnamefont {S.}~\bibnamefont {Jin}}, \bibinfo {author} {\bibfnamefont
  {Z.}~\bibnamefont {Xu}},\ and\ \bibinfo {author} {\bibfnamefont
  {L.}~\bibnamefont {Li}},\ }\href {https://doi.org/10.1063/5.0107140}
  {\bibfield  {journal} {\bibinfo  {journal} {The Journal of Chemical Physics}\
  }\textbf {\bibinfo {volume} {157}},\ \bibinfo {pages} {144102} (\bibinfo
  {year} {2022})}\BibitemShut {NoStop}%
\bibitem [{\citenamefont {Tsai}(1979)}]{Tsai:JChemPhysL:1979}%
  \BibitemOpen
  \bibfield  {author} {\bibinfo {author} {\bibfnamefont {D.~H.}\ \bibnamefont
  {Tsai}},\ }\href {https://doi.org/10.1063/1.437577} {\bibfield  {journal}
  {\bibinfo  {journal} {The Journal of Chemical Physics}\ }\textbf {\bibinfo
  {volume} {70}},\ \bibinfo {pages} {1375} (\bibinfo {year}
  {1979})}\BibitemShut {NoStop}%
\bibitem [{\citenamefont {Goldstein}(1980)}]{goldstein:mechanics}%
  \BibitemOpen
  \bibfield  {author} {\bibinfo {author} {\bibfnamefont {H.}~\bibnamefont
  {Goldstein}},\ }\href@noop {} {\emph {\bibinfo {title} {Classical
  Mechanics}}}\ (\bibinfo  {publisher} {Addison-Wesley},\ \bibinfo {year}
  {1980})\BibitemShut {NoStop}%
\bibitem [{\citenamefont {Park}(1990)}]{Park1990}%
  \BibitemOpen
  \bibfield  {author} {\bibinfo {author} {\bibfnamefont {D.}~\bibnamefont
  {Park}},\ }\bibinfo {title} {N-particle systems},\ in\ \href
  {https://doi.org/10.1007/978-3-642-74922-3_4} {\emph {\bibinfo {booktitle}
  {Classical Dynamics and Its Quantum Analogues}}}\ (\bibinfo  {publisher}
  {Springer Berlin Heidelberg},\ \bibinfo {address} {Berlin, Heidelberg},\
  \bibinfo {year} {1990})\ pp.\ \bibinfo {pages} {82--102}\BibitemShut
  {NoStop}%
\bibitem [{\citenamefont {Caillol}(1999)}]{Caillol:JChemPhys:1999}%
  \BibitemOpen
  \bibfield  {author} {\bibinfo {author} {\bibfnamefont {J.~M.}\ \bibnamefont
  {Caillol}},\ }\href {https://doi.org/10.1063/1.479965} {\bibfield  {journal}
  {\bibinfo  {journal} {The Journal of Chemical Physics}\ }\textbf {\bibinfo
  {volume} {111}},\ \bibinfo {pages} {6538} (\bibinfo {year}
  {1999})}\BibitemShut {NoStop}%
\bibitem [{\citenamefont {Hansen}(1973)}]{Hansen:PRA:1973}%
  \BibitemOpen
  \bibfield  {author} {\bibinfo {author} {\bibfnamefont {J.~P.}\ \bibnamefont
  {Hansen}},\ }\href {https://doi.org/10.1103/PhysRevA.8.3096} {\bibfield
  {journal} {\bibinfo  {journal} {Phys. Rev. A}\ }\textbf {\bibinfo {volume}
  {8}},\ \bibinfo {pages} {3096} (\bibinfo {year} {1973})}\BibitemShut
  {NoStop}%
\bibitem [{\citenamefont {de~Leeuw}\ \emph {et~al.}(1980)\citenamefont
  {de~Leeuw}, \citenamefont {Perram},\ and\ \citenamefont
  {Smith}}]{DeLeeuw1980SimulationOE}%
  \BibitemOpen
  \bibfield  {author} {\bibinfo {author} {\bibfnamefont {S.~W.}\ \bibnamefont
  {de~Leeuw}}, \bibinfo {author} {\bibfnamefont {J.~W.}\ \bibnamefont
  {Perram}},\ and\ \bibinfo {author} {\bibfnamefont {E.~R.}\ \bibnamefont
  {Smith}},\ }\href {https://doi.org/https://doi.org/10.1098/rspa.1980.0135}
  {\bibfield  {journal} {\bibinfo  {journal} {Proceedings of the Royal Society
  of London. A. Mathematical and Physical Sciences}\ }\textbf {\bibinfo
  {volume} {373}},\ \bibinfo {pages} {27} (\bibinfo {year} {1980})}\BibitemShut
  {NoStop}%
\bibitem [{\citenamefont {Fukuda}\ and\ \citenamefont
  {Nakamura}(2022)}]{Fukuda2022}%
  \BibitemOpen
  \bibfield  {author} {\bibinfo {author} {\bibfnamefont {I.}~\bibnamefont
  {Fukuda}}\ and\ \bibinfo {author} {\bibfnamefont {H.}~\bibnamefont
  {Nakamura}},\ }\href {https://doi.org/10.1007/s12551-022-01029-2} {\bibfield
  {journal} {\bibinfo  {journal} {Biophysical Reviews}\ }\textbf {\bibinfo
  {volume} {14}},\ \bibinfo {pages} {1315} (\bibinfo {year}
  {2022})}\BibitemShut {NoStop}%
\bibitem [{\citenamefont {Jha}\ \emph {et~al.}(2010)\citenamefont {Jha},
  \citenamefont {Sknepnek}, \citenamefont {Guerrero-Garc{\'i}a},\ and\
  \citenamefont {Olvera de~la Cruz}}]{Jha:2010}%
  \BibitemOpen
  \bibfield  {author} {\bibinfo {author} {\bibfnamefont {P.~K.}\ \bibnamefont
  {Jha}}, \bibinfo {author} {\bibfnamefont {R.}~\bibnamefont {Sknepnek}},
  \bibinfo {author} {\bibfnamefont {G.~I.}\ \bibnamefont
  {Guerrero-Garc{\'i}a}},\ and\ \bibinfo {author} {\bibfnamefont
  {M.}~\bibnamefont {Olvera de~la Cruz}},\ }\href
  {https://doi.org/10.1021/ct100365c} {\bibfield  {journal} {\bibinfo
  {journal} {Journal of Chemical Theory and Computation}\ }\textbf {\bibinfo
  {volume} {6}},\ \bibinfo {pages} {3058} (\bibinfo {year} {2010})}\BibitemShut
  {NoStop}%
\bibitem [{\citenamefont {Jin}\ \emph {et~al.}(2021)\citenamefont {Jin},
  \citenamefont {Li}, \citenamefont {Xu},\ and\ \citenamefont
  {Zhao}}]{Correct_grad}%
  \BibitemOpen
  \bibfield  {author} {\bibinfo {author} {\bibfnamefont {S.}~\bibnamefont
  {Jin}}, \bibinfo {author} {\bibfnamefont {L.}~\bibnamefont {Li}}, \bibinfo
  {author} {\bibfnamefont {Z.}~\bibnamefont {Xu}},\ and\ \bibinfo {author}
  {\bibfnamefont {Y.}~\bibnamefont {Zhao}},\ }\href
  {https://doi.org/10.1137/20M1371385} {\bibfield  {journal} {\bibinfo
  {journal} {SIAM Journal on Scientific Computing}\ }\textbf {\bibinfo {volume}
  {43}},\ \bibinfo {pages} {B937} (\bibinfo {year} {2021})}\BibitemShut
  {NoStop}%
\bibitem [{\citenamefont {Hamaguchi}\ and\ \citenamefont
  {Farouki}(1994)}]{Hamaguchi:JChemPhys:1994}%
  \BibitemOpen
  \bibfield  {author} {\bibinfo {author} {\bibfnamefont {S.}~\bibnamefont
  {Hamaguchi}}\ and\ \bibinfo {author} {\bibfnamefont {R.~T.}\ \bibnamefont
  {Farouki}},\ }\href {https://doi.org/10.1063/1.467954} {\bibfield  {journal}
  {\bibinfo  {journal} {The Journal of Chemical Physics}\ }\textbf {\bibinfo
  {volume} {101}},\ \bibinfo {pages} {9876} (\bibinfo {year}
  {1994})}\BibitemShut {NoStop}%
\bibitem [{\citenamefont {Fraser}\ \emph {et~al.}(1996)\citenamefont {Fraser},
  \citenamefont {Foulkes}, \citenamefont {Rajagopal}, \citenamefont {Needs},
  \citenamefont {Kenny},\ and\ \citenamefont {Williamson}}]{Fraser:PRB:1996}%
  \BibitemOpen
  \bibfield  {author} {\bibinfo {author} {\bibfnamefont {L.~M.}\ \bibnamefont
  {Fraser}}, \bibinfo {author} {\bibfnamefont {W.~M.~C.}\ \bibnamefont
  {Foulkes}}, \bibinfo {author} {\bibfnamefont {G.}~\bibnamefont {Rajagopal}},
  \bibinfo {author} {\bibfnamefont {R.~J.}\ \bibnamefont {Needs}}, \bibinfo
  {author} {\bibfnamefont {S.~D.}\ \bibnamefont {Kenny}},\ and\ \bibinfo
  {author} {\bibfnamefont {A.~J.}\ \bibnamefont {Williamson}},\ }\href
  {https://doi.org/10.1103/PhysRevB.53.1814} {\bibfield  {journal} {\bibinfo
  {journal} {Phys. Rev. B}\ }\textbf {\bibinfo {volume} {53}},\ \bibinfo
  {pages} {1814} (\bibinfo {year} {1996})}\BibitemShut {NoStop}%
\end{thebibliography}

\providecommand{\noopsort}[1]{}\providecommand{\singleletter}[1]{#1}%

\end{document}